\documentclass[12pt]{iopart}

\usepackage{color}
\usepackage[utf8]{inputenc}
\usepackage{graphicx}

\usepackage{etoolbox}
\makeatletter
\def\@mkboth#1#2{}
\newlength\appendixwidth
\preto\appendix{\addtocontents{toc}{\protect\patchl@section}}
\newcommand{\patchl@section}{%
  \settowidth{\appendixwidth}{\textbf{Appendix }}%
  \addtolength{\appendixwidth}{1.5em}%
  \patchcmd{\l@section}{1.5em}{\appendixwidth}{}{\ddt}%
}
\makeatother

\newcommand{\nuc}[2]{$^{#1}$#2}

\definecolor{RED}{RGB}{255, 0, 0}

\date{\today}

\begin{document}

\title[Nuclear Isomers in the Era of FRIB]{Executive Summary of the Topical Program: \\
Nuclear Isomers in the Era of FRIB\footnote{This document is intended for unlimited release under LA-UR-23-22994.}\footnote{ARL Distribution Statement A.  Approved for public release: distribution unlimited.}}

\newcommand{\lanl}{$^{1}$}
\newcommand{\anl}{$^{2}$}
\newcommand{\lsu}{$^{3}$}
\newcommand{\fsu}{$^{4}$}
\newcommand{\nd}{$^{5}$}
\newcommand{\msu}{$^{6}$}
\newcommand{\frib}{$^{7}$}
\newcommand{\arl}{$^{8}$}
\newcommand{\ornl}{$^{9}$}
\newcommand{\utk}{$^{10}$}
\newcommand{\msstate}{$^{11}$}
\newcommand{\llnl}{$^{12}$}
\newcommand{\gsi}{$^{13}$}
\newcommand{\clemson}{$^{14}$}
\newcommand{\lbnl}{$^{15}$}
\newcommand{\sjtu}{$^{16}$}
\newcommand{\surrey}{$^{17}$}

\newcommand{\presinst}{$^{99}$}

\author{
  G.~W.~Misch\lanl,
  M.~R.~Mumpower\lanl,
  F.~G.~Kondev\anl,
  S.~T.~Marley\lsu,
  S.~Almaraz-Calderon\fsu,
  M.~Brodeur\nd,
  B.~A.~Brown\msu$^,$\frib,
  M.~P.~Carpenter\anl,
  J.~J.~Carroll\arl,
  C.~J.~Chiara\arl,
  K.~A.~Chipps\ornl$^,$\utk,
  B.~P.~Crider\msstate,
  A.~Gade\msu$^,$\frib,
  R.~Grzywacz\utk,
  K.~L.~Jones\utk,
  B.~P.~Kay\anl,
  K.~Kolos\llnl,
  Yu.~A.~Litvinov\gsi,
  S.~Lopez-Caceres\lsu,
  B.~S.~Meyer\clemson,
  K.~Minamisono\msu$^,$\frib,
  G.~E.~Morgan\lsu$^,$\anl,
  R.~Orford\lbnl,
  S.~D.~Pain\ornl$^,$\utk,
  J.~Purcell\msu,
  A.~Ratkiewicz\llnl,
  H.~Schatz\msu$^,$\frib,
  T.~M.~Sprouse\lanl,
  Y.~Sun{\sjtu},
  R.~Surman\nd,
  J.~A.~Tannous{\clemson},
  P.~M.~Walker\surrey
}

\address{{\lanl}Los Alamos National Laboratory}
\address{{\anl}Argonne National Laboratory}
\address{{\lsu}Louisiana State University}
\address{{\fsu}Florida State University}
\address{{\nd}University of Notre Dame}
\address{{\msu}Michigan State University}
\address{{\frib}Facility for Rare Isotope Beams}
\address{{\arl}US Army Combat Capabilities Development Command Army Research Laboratory}
\address{{\ornl}Oak Ridge National Laboratory}
\address{{\utk}University of Tennessee Knoxville}
\address{{\msstate}Mississippi State University}
\address{{\llnl}Lawrence Livermore National Laboratory}
\address{{\gsi}GSI Helmholtzzentrum f{\"u}r Schwerionenforschung GmbH}
\address{{\clemson}Clemson University}
\address{{\lbnl}Lawrence Berkeley National Laboratory}
\address{{\sjtu}Shanghai Jiao Tong University}
\address{{\surrey}University of Surrey}

\eads{
  \mailto{wendell@lanl.gov},
  \mailto{mumpower@lanl.gov},
  \mailto{kondev@anl.gov},
  \mailto{smarley@lsu.edu}
}

\maketitle

We report on the workshop ``Nuclear Isomers in the Era of FRIB'', held at the Facility for Rare Isotope Beams (FRIB) at Michigan State University, May 9--20, 2022.
There were 32 participants, 27 formal presentations (including 4 from graduate students and postdocs) representing institutions from the US, the UK, Germany, and China.  
The presentations and numerous discussions covered past and present efforts, both experimental and theoretical, to study and characterize nuclear isomers. 
The meeting clearly demonstrated that isomer studies enable nuclear science and feature prominently in all four FRIB science pillars: properties of rare isotopes, nuclear astrophysics, fundamental symmetries, and applications for the nation and society.

Nuclear isomers are metastable states with long half lives ($t_{1/2} \geq 10^{-9} s$) \cite{walker1999energy} compared to typical nuclear excited states ($t_{1/2} \sim 10^{-12} s$).
In addition to electromagnetic transitions, isomers may decay via any known weak or strong process (e.g., $\beta^\pm$, baryonic particle emission, spontaneous fission, etc.).
Some isotopes contain isomers more stable than their ground states; the most notable is the longest-lived known isomer \nuc{180m}{Ta} ($t_{1/2} > 10^{16} y$ \cite{lehnert2017search}) that is essentially stable compared to the 8.15-hour \nuc{180}Ta ground state.
Isotopes can contain multiple isomeric states, and they can have excitation energies ranging from a few eV (\nuc{229m}{Th} \cite{sikorsky2020measurement}) to over 10 MeV (\nuc{208}{Pb} \cite{broda2017doubly}, \nuc{152}{Er} \cite{kuhnert1992observation}), well into the particle-decay continuum.

\vspace{5pt}
\noindent\textbf{Nuclear Properties}

The relative stability of isomers arises from radical differences between their quantum properties and those of lower-lying states.
These properties include spin ($J$), projection of spin along an axis of symmetry ($K$), shape, and seniority (number of broken nucleon pairs) \cite{walker2020100}.
This diversity of isomers makes them useful tools to test and refine our understanding of nuclei across the nuclear landscape.
For instance, microsecond isomers live long enough to be transported and studied in a low-background environment.
Intense isomeric beams accelerated above Coulomb-barrier energies can be unique probes of nuclear structure in direct reaction studies \cite{santiagogonzalez2018probing}.
Some isomers illuminate coupling of nuclear and atomic degrees of freedom \cite{chiara2018isomer}.

Modeling isomers can be challenging.
The nuclear shell model enables predictions of spin traps and seniority isomers, while the liquid drop model can explain shape isomers and the related $K$ isomers.
However, to better understand future data, new models must be developed.
These theoretical guides lead to better targeted experiments, which in turn feed back into improved theoretical models and better predictions where data are limited.
With the era of FRIB comes an unprecedented opportunity for isomer exploration and discovery.

\vspace{5pt}
\noindent\textbf{Nuclear Astrophysics}

Isomeric beams offer the unique opportunity to experimentally determine nuclear transmutation rates of thermally populated excited states in hot astrophysical environments.
Moreover, isomers that remain metastable in astrophysical environments (astromers \cite{misch2020astromers}) can affect energy release, nucleosynthesis, and electromagnetic signals that complement multi-messenger observations.
Isomers in different isotopes are thermally populated or manifest as astromers under different conditions, so detailed studies of particular species are required for distinct environments.

The diverse production and destruction mechanisms of astromers and other isomers necessitate a range of relevant nuclear data, including excitation energies, spin/parity assignments, decay branching ratios, and properties of nearby excited states.
Therefore, an array of experimental techniques are required to incorporate isomers' behavior into astrophysical transmutation rate databases for use in modeling.

\vspace{5pt}
\noindent\textbf{Fundamental Symmetries}

Isomers' metastable character affords exceptional opportunities to study the foundations of our physical universe.
Isomers with a strong octopole moment and opposite parity to their ground state are sensitive to charge-parity reversal (CP) violation, and hence time-reversal (T) violation \cite{singh2019pa229}.
Candidate isotopes have been identified that await discovery of their isomers.
Isomers have also been proposed for searches of weakly interacting dark matter and topological defect dark matter. 
For example, if a dark matter particle induces a transition from a long-lived isomer to a nearby excited state, subsequent decay radiation could be used as a detection signature.

\vspace{5pt}
\noindent\textbf{Applications for Society}

An isomer with a suitably long half-life, decay energy, and economical production mechanism could provide the necessary ingredients for novel applications.
Currently, the most impactful are in nuclear medicine applications such as imaging (\nuc{99m}{Tc}) and cancer treatment.
Growing experimental capabilities and efforts provide immense opportunities not only for harvesting medical isomers, but also for identifying and characterizing potential new treatments.

Additionally, long-lived isomers that can be depopulated on demand would be the essential component in high-energy-density nuclear batteries \cite{carroll2018nuclear}, and ongoing work aims to open the door to $\gamma$-ray lasers \cite{baldwin1981approaches}.
The low-energy isomer in \nuc{229}{Th} is a prime candidate for a nuclear clock that is robust against environmental influence and with 1--2 orders of magnitude greater precision than the best atomic clocks \cite{campbell2012single}.
 
\vspace{12pt}
\noindent\textbf{Resolution}

Nuclear isomers are worth studying in their own right because of their far-reaching impact: they are important aspects of nuclear structure that are difficult to model precisely using current theoretical approaches, they influence astrophysical processes, they can be probes of fundamental symmetries, and they have practical applications in technology.

The new generation of radioactive ion beam facilities---including FRIB---create enormous opportunities for the discovery, characterization, and use of isomers.
We recommend that facilities enable and the community prioritize isomer research, both as targeted experiments and as experiments of opportunity.

\vspace{12pt}
\noindent\textbf{Acknowledgment}

We thank the FRIB Theory Alliance for funding and hosting this topical program.

\pagebreak
\noindent\textbf{References}
\bibliographystyle{unsrt}
\bibliography{bibliography}

\begin{thebibliography}{10}

\bibitem{walker1999energy}
Philip Walker and George Dracoulis.
\newblock Energy traps in atomic nuclei.
\newblock {\em Nature}, 399(6731):35--40, 1999.

\bibitem{lehnert2017search}
Bj{\"o}rn Lehnert, Mikael Hult, Guillaume Lutter, and Kai Zuber.
\newblock Search for the decay of nature's rarest isotope $^{180m}\mathrm{Ta}$.
\newblock {\em Physical Review C}, 95(4):044306, 2017.

\bibitem{sikorsky2020measurement}
Tomas Sikorsky, Jeschua Geist, Daniel Hengstler, Sebastian Kempf, Loredana
  Gastaldo, Christian Enss, Christoph Mokry, J\"org Runke, Christoph~E.
  D\"ullmann, Peter Wobrauschek, Kjeld Beeks, Veronika Rosecker, Johannes~H.
  Sterba, Georgy Kazakov, Thorsten Schumm, and Andreas Fleischmann.
\newblock Measurement of the $^{229}\mathrm{Th}$ isomer energy with a magnetic
  microcalorimeter.
\newblock {\em Physical Review Letters}, 125:142503, 2020.

\bibitem{broda2017doubly}
R~Broda, RVF Janssens, {\L}W~Iskra, J~Wrzesinski, B~Fornal, MP~Carpenter,
  CJ~Chiara, N~Cieplicka-Ory{\'n}czak, CR~Hoffman, FG~Kondev, et~al.
\newblock Doubly magic $^{208}$\uppercase{P}b: High-spin states, isomers, and
  $\uppercase{E}3$ collectivity in the yrast decay.
\newblock {\em Physical Review C}, 95(6):064308, 2017.

\bibitem{kuhnert1992observation}
A~Kuhnert, D~Alber, H~Grawe, H~Kluge, KH~Maier, W~Reviol, X~Sun, EM~Beck,
  AP~Byrne, H~H{\"u}bel, et~al.
\newblock Observation of high-spin states in the $\uppercase{N}=84$ nucleus
  $^{152}$\uppercase{E}r and comparison with shell-model calculations.
\newblock {\em Physical Review C}, 46(2):484, 1992.

\bibitem{walker2020100}
Philip Walker and Zsolt Podoly{\'a}k.
\newblock 100 years of nuclear isomers—then and now.
\newblock {\em Physica Scripta}, 95(4):044004, 2020.

\bibitem{santiagogonzalez2018probing}
D.~Santiago-Gonzalez, K.~Auranen, M.~L. Avila, A.~D. Ayangeakaa, B.~B. Back,
  S.~Bottoni, M.~P. Carpenter, J.~Chen, C.~M. Deibel, A.~A. Hood, C.~R.
  Hoffman, R.~V.~F. Janssens, C.~L. Jiang, B.~P. Kay, S.~A. Kuvin, A.~Lauer,
  J.~P. Schiffer, J.~Sethi, R.~Talwar, I.~Wiedenh\"over, J.~Winkelbauer, and
  S.~Zhu.
\newblock Probing the single-particle character of rotational states in
  $^{19}\mathrm{F}$ using a short-lived isomeric beam.
\newblock {\em Physical Review Letters}, 120:122503, 2018.

\bibitem{chiara2018isomer}
CJ~Chiara, JJ~Carroll, MP~Carpenter, JP~Greene, DJ~Hartley, RVF Janssens,
  GJ~Lane, JC~Marsh, DA~Matters, M~Polasik, et~al.
\newblock Isomer depletion as experimental evidence of nuclear excitation by
  electron capture.
\newblock {\em Nature}, 554(7691):216--218, 2018.

\bibitem{misch2020astromers}
G~Wendell Misch, Surja~K Ghorui, Projjwal Banerjee, Yang Sun, and Matthew~R
  Mumpower.
\newblock Astromers: nuclear isomers in astrophysics.
\newblock {\em The Astrophysical Journal Supplement Series}, 252(1):2, 2020.

\bibitem{singh2019pa229}
Jaideep~Taggart Singh.
\newblock A new concept for searching for time-reversal symmetry violation
  using {Pa}-229 ions trapped in optical crystals.
\newblock {\em Hyperfine Interactions}, 240(1):29, 2019.

\bibitem{carroll2018nuclear}
James~J Carroll.
\newblock Nuclear metastables for energy and power: status and challenges.
\newblock {\em Innovations in Army Energy and Power Materials Technologies},
  36:289, 2018.

\bibitem{baldwin1981approaches}
George~C Baldwin, Johndale~C Solem, and Vitalii~I Gol'Danskii.
\newblock Approaches to the development of gamma-ray lasers.
\newblock {\em Reviews of Modern Physics}, 53(4):687, 1981.

\bibitem{campbell2012single}
Corey~J Campbell, Alexander~G Radnaev, A~Kuzmich, Vladimir~A Dzuba, Victor~V
  Flambaum, and Andrei Derevianko.
\newblock Single-ion nuclear clock for metrology at the 19th decimal place.
\newblock {\em Physical Review Letters}, 108(12):120802, 2012.

\end{thebibliography}

\end{document}